# Automatic quality evaluation and (semi-) automatic improvement of OCR models for historical printings

Uwe Springmann · Florian Fink · Klaus U. Schulz

**Abstract** Good OCR results for historical printings rely on the availability of recognition models trained on diplomatic transcriptions as ground truth, which is both a scarce resource and time-consuming to generate. Instead of having to train a separate model for each historical typeface, we propose a strategy to start from models trained on a combined set of available transcriptions from 6 printings ranging from 1471 to 1686 in a variety of fonts. These *mixed models* result in character accuracy rates (defined as the ratio of correctly recognized characters to the total number of characters in the OCR output) over 90% on a test set of another 6 printings from the same period of time, but without any representation in the training data, demonstrating the possibility to overcome the typography barrier by generalizing from a few typefaces to a larger set of (similar) fonts in use over a period of time. The output of these mixed models is then used as a baseline to be further improved by both fully automatic methods (taking the OCR result of mixed models as pseudo ground truth for subsequent training) and semi-automatic methods involving a minimal amount of manual transcriptions.

In order to evaluate the recognition quality of each model in a series of models generated during the training process in the absence of any ground truth, we introduce two readily observable quantities that correlate well with true accuracy, giving us an ordinal ranking scale which allows to automatically select the (nearly) best performing model for recognition. These quantities are *mean character confidence C* (as given by the OCR engine OCRopus) and *mean token lexicality L* (a distance measure of OCR tokens from modern wordforms taking historical spelling patterns into account, which can be calculated for any OCR engine). Whereas the fully automatic method is able to improve upon the result of a mixed model by only 1-2 percentage points, already 100-200 hand-corrected lines lead to much better OCR results with character error rates of only a few percent. This procedure minimizes the amount of ground truth production and does not depend on the previous construction of a specific typographic model.

**CR Subject Classification** I.7.5

This work was partially funded by Deutsche Forschungsgemeinschaft (DFG) under grant no. SCHU-1026/7-1 and LU 856/7-1.

Uwe Springmann
Humboldt-Universität zu Berlin

Uwe Springmann · Florian Fink · Klaus U. Schulz
Ludwig-Maximilians-Universität München
E-mail: (springmann, finkf, schulz) @ cis.uni-muenchen.de

# 1 Introduction

In the last years several attempts were made to develop OCR methods for historical documents. As a lot of the early printings have already been digitized (in the sense of making scanned images available), the bottleneck for getting access to the contents of these books now consists in methods of conversion of these images to electronic, machine-actionable text data. Commercial OCR engines lack the possibility to get trained on early typographies and give unsatisfactory results of at most 85% character accuracy on early printings (Reddy and Crane 2006; Piotrowski 2012; Strange et al 2014; Springmann et al 2014), and are deemed completely useless for very early printings (incunabula printings before 1501; Rydberg-Cox 2009).

OCRopus with its new recognizer based upon a recurrent neural network with LSTM architecture has been shown to be trainable on historic fonts and deliver competitive or even better results than either Tesseract or ABBYY on 18th and 19th century Fraktur printings (Breuel et al 2013). These results were achieved by generating a lot of training material automatically, generating artificially degraded images from existing text and computer fonts. This method does not work very well for very early printings (Springmann et al 2014), probably because we lack computer fonts that are similar enough to the actual printings and because the interword spacings are highly irregular, leading to OCR tokens being merged to a single long string without any intervening spaces. Training on real images solves both of these problems (Springmann and Lüdeling 2016) but requires the time consuming task of diplomatic transcription of the training material. While the resulting OCR model works very well for the book it has been trained on (often reaching accuracies of 98% for even the earliest printings), these individual models do not generalize well to other books. In an automatic setting where a large amount of books need to be OCRed in short time, the training of individual models is out of the question. The construction of mixed models trained on material from several different books partly overcomes this problem with accuracies still over 90% for a wide variety of books (Springmann and Lüdeling 2016), but in the absence of ground truth to test against one cannot know just how good the mixed model is for a particular book or how one could decide whether one model is better than another one without manually counting the errors in the OCR result. An optimal strategy seems to be to start from a mixed model and to refine it later (either automatically or with minimal manual effort), which requires a measure for accuracy independent of ground truth.

The goal of this paper is to answer the following questions:

1. In the absence of ground truth, how can the recognition quality of OCRopus (or any other OCR engine) be estimated automatically?
2. Given an automatic method for OCR quality estimation, how could one use it to construct a model better than the start model *in a fully automatic way*, and what improvement can be expected?
3. What OCR quality can be obtained by adding a small amount of manual work, preparing some lines of ground truth? What is the tradeoff between the number of training lines and the model improvement? How good can it get compared to an individual model trained on a large set of ground truth data from the same document?

Question 1 is of great importance for any large digitization program, where one wants to get a quick estimate of the OCR quality of a book, a page, or a paragraph. To summarize our results:

1. Both *mean token lexicality L*, a profiler-based measure explained below, and *mean character confidence C* calculated from individual character confidences given by the OCRopus engine (see Fig. 1), correlate well with character accuracy. Measurement of either quantity therefore leads to an estimate of true accuracy. While we find a tighter correlation of accuracy with confidence, lexicality can be calculated for any OCR engine, regardless of the quality of confidence values in the output.
2. For all documents in our data set, using the recognition result of a mixed model as a starting point, appropriate forms of fully automatic training result in an improved quality of a few percentage points in accuracy. For model selection after training, the above techniques for estimating accuracy are essential.
3. The manual correction of as few as 100 - 200 text lines and subsequent training on just these lines often leads to excellent models. Even good starting models can be considerably improved. As to the selection of lines, a mixture of randomly selected lines together with a set of poorly recognized lines seems to have the best effect. In general, adding more training material leads to better results with diminishing returns, approaching the accuracy of an individual model trained on large amounts of ground truth.

The paper is organized as follows: Section 2 gives a short account of the state of the art for OCR of historical documents to put our work in perspective. Section 3 describes the data sets for our models and experiments. Section 4 describes lexicality and mean character confidence and shows their correlation with accuracy. The following sections report our experiments and their outcomes for the automatic (Section 5) and semi-automatic (Section 6) method, and we end with a summary (Section 7).

## 2 Related work

Work by other groups has mostly focused on Tesseract, which is trainable on artificial images generated from computer fonts in a similar way as OCRopus. Training on real data, however, has proved to be difficult, and lead to efforts to reconstruct the original typeset from cut-out glyphs. This has been done by both the Digital Libraries team of the Poznań Supercomputing and Networking Center (Dudczak et al 2014) with their cutouts application[1] (proprietary) and EMOP's Franken+ tool[2] (open source). However, the latter group has reported on reaching only about 86% accuracy on the ECCO document collection and 68% on the EEBO collection.[3] Their OCR suffers badly from scans of binarized microfilm images containing a lot of noise. A publication from this project with a title similar to the present one (Gupta et al 2015) consequently deals with improving OCR quality by automat-ically distinguisting between text and non-text areas. The combined models published from this project covering a variety of typesets (similar to our mixed models) do not presently give high accuracies above 90%.

---

[1] https://confluence.man.poznan.pl/community/display/WLT/Cutouts+application
[2] http://emop.tamu.edu/outcomes/Franken-Plus
[3] http://emop.tamu.edu/final-report

The Kallimachos project[4] at Würzburg University did have success with Franken+ to reach accuracies over 95% for an incunable printing (Kirchner et al 2016) but this method relies again on creating diplomatic transcriptions from scratch for each individual typeface. The method proposed by Ul-Hasan et al (2016) to circumvent ground truth production by first training Tesseract on a historically reconstructed typeface with subsequent OCRopus training on the actual book using Tesseract's recognition as pseudo ground truth has also achieved accuracies above 95% but shifts the transcription effort to the manual (re)construction of the typeface.

A completely different approach was taken with the new Ocular OCR engine by Berg-Kirkpatrick et al (2013) which is able to convert historical printings to electronic text in a completely unsupervised manner (i.e., no ground truth is needed) employing a language, typesetting, inking and noise model. This may be a viable alternative for training individual models with low manual effort, but it seems to be very resource-intensive and slow (the recognition of 30 printed lines takes 2.4 minutes according to Berg-Kirkpatrick and Klein 2014). Its results are better than (untrained) Tesseract and ABBYY, but it remains to be shown that this method is able to consistently reach accuracies higher than 90%.

In summary, while there are other approaches to train individual OCR models for the recognition of historical documents, none have so far reported results as good as OCRopus (consistently over 95% accuracy), nor has it been shown that one could construct generalized models applicable to a variety of books with reasonable results (above 90% accuracy).

## 3 Data sets for training and evaluation - mixed standard models

The data sets used for training and testing our individual and mixed models consist of twelve Latin books printed with Antiqua types from 1471 to 1686. We deliberately chose these early printings, among them four incunabula printed before 1501, because no other OCR methods have been able to yield character accuracies consistently over 95% for such material (see Section 1). Scans for these books have been downloaded from archive.org[5] and the Bavarian State Library.[6] The training and testing data consist of a set of printed line images extracted from book pages together with their diplomatic transcriptions serving as ground truth for model training and for the evaluation of the recognition error rate. These data have been manually compiled over the course of the last two years by one of the authors (US) with some help by students and colleagues. Table 1 gives bibliographic details on these books as well as the amount of data (number of lines) available. We split our material in two parts covering almost the same range of printing years: Within the chronological order of all books, the even-numbered books make up one part, the odd-numbered books the other part. A mixed model was trained on each part and applied to both the books of its own part (representing the case where the specific typesets of these books contributed to the model's training set) and to the books of the other part (where these books had no representation in the training set). The first case corresponds to a kind of omnifont model, but given the high number of individual typesets that were in use historically, it is presently infeasible to construct such a model for any specific period except as a toy model for comparison purposes. Until a large number of transcriptions covering the history of modern printing become available for training, the second case of a mixed model applied to hitherto unseen typesets will be the default starting point (except if the document to be OCRed was printed in a typeface that happened to be in the training set). The evaluation of results applying the mixed (other) models gives an indication of how well mixed models generalize to hitherto unseen typesets, and whenever we speak of mixed models in general in the remainder of this paper we exactly mean these mixed (other) models.

---

[4] `kallimachos.de`
[5] `http://www.archive.org`
[6] `http://www.digitale-sammlungen.de/index.html?&l=en`

**Table 1** Data sets from 12 Latin books separated in two parts. Given are the printing year, short title, author, the number of available lines with ground truth for training and evaluation and a label (year plus the first letter of the short title) by which these books get referred to in the text.

| Year | (Short) Title | Author | # lines | label |
|---|---|---|---|---|
| *Part 1* | | | | |
| 1476 | Speculum Naturale | Beauvais | 2012 | 1476-S |
| 1497 | Stultifera Navis | Brant/Locher | 1092 | 1497-S |
| 1543 | De Bello Alexandrino | Caesar | 832 | 1543-D |
| 1553 | Carmina | Pigna | 298 | 1553-C |
| 1557 | Methodus | Clenardus | 350 | 1557-M |
| 1686 | Lexicon Atriale | Comenius | 1105 | 1686-L |
| *Part 2* | | | | |
| 1471 | Orthographia | Tortellius | 417 | 1471-O |
| 1483 | Decades | Biondo | 915 | 1483-D |
| 1522 | De Septem Secundadeis | Trithemius | 201 | 1522-D |
| 1564 | Thucydides | Valla | 1948 | 1564-T |
| 1591 | Progymnasmata vol. I | Pontanus | 710 | 1591-P |
| 1668 | Leviathan | Hobbes | 1078 | 1668-L |

For training an individual model, about 20% of the available lines for each book have been set aside to form the test set, and the remainder was used as training material (the division was done on a pagewise basis). The resulting model was saved every 1,000 learning steps, where each step con-sists in seeing one line image and its associated ground truth. After training for some thousand steps, the model with the best accuracy on the test set was chosen for later recognition tasks. The results of these and all other training experiments are shown in Fig. 6.

The mixed models were trained by pooling the training sets of all the books belonging to Part 1 and Part 2, respectively. For good recognition results of a mixed model it is necessary to standardize the transcriptions (probably originating from various sources with different transcription guidelines, both explicit written guidelines and unconscious ones) of all the books contributing to the pooled training set so that the same glyph is coded with the same Unicode code point.

The character accuracies of OCR text from books in one part resulting from the application of the mixed model trained on the other part are shown as yellow columns in Fig. 6 and mostly have accuracies over 90% (with two exceptions). This shows that mixed models generalize fairly well over a range of typographies. This is not at all true for individual models, which have high error rates when applied to other books even if the typeset looks very similar to the human eye (Springmann and Lüdeling 2016). The mixed (other) model can therefore be taken as a starting point for subsequent model improvements.

The difference in accuracy in Fig. 6 between the mixed (other) model and the individual models may be seen as the improvement potential, which we try to partially realize with our experiments to train new individual models with no (fully automatic) or minimal (semi-automatic) manual effort.

## 4 Automatic quality evaluation

Let us assume that we have a set of mixed models at our disposal and we want to select the model generating the best OCR result for a document as the starting point for further improvements. However, in a realistic scenario we will apply a model to a document for which we do not have any ground truth. How can we know if the model works well? This is a core problem in large digitization projects where thousands of books are processed, each having specific problems. The question arises: *How can we estimate OCR quality in the absence of any ground truth data?*

Methods for automatically testing OCR quality are therefore important both for quality control of the OCR result from a single model and for the selection of the most appropriate model with respect to a given printed document. This latter situation is considered below. Furthermore, methods for testing OCR quality can be used to obtain a kind of diagnostics when OCR results for a book do not meet the expectations. Quality testing helps to find those subparts (pages and lines) where serious problems arise, and to find hints on how to improve a model (cf. Figure 3). In a similar way, OCR quality estimates indicate if there is potential to improve a model, and they can be used to guide the selection of lines to be used as ground truth in model training (see Sect. 6).

Because OCR quality cannot be directly observed, we need a kind of substitute that is easily measured and correlates well with accuracy. We propose two such approximate measures for OCR accuracy: One is the *lexicality* of OCR tokens determined by our language-aware OCR error profiler (Reffle and Ringlstetter 2013). For each OCR token the profiler calculates the minimum edit distance (Levenshtein distance) to its most probable modern lexical equivalent, discounting any differences due to historical spelling patterns. The printed word *judicare*, recognized as *judicarc* and with a modern equivalent *iudicare*, therefore has a Levenshtein distance of 1 (OCR error: $e \to c$), as the historical spelling pattern $i \to j$ does not get counted. The sum of these Levenshtein distances over all tokens is therefore a (statistical) measure for the OCR errors of the text, and the lexicality defined as *L = (1 - mean Levenshtein distance per character)* is a measure for accuracy. Problems with this measure arise from lexical gaps (mostly proper names) and very garbled tokens (either too short such as sequences of single letters, or too long because of merged tokens with unrecognized whitespaces) which do not get Levenshtein distances assigned.

The other measure are the confidence values that OCRopus assigns to its output characters[7]. Whenever an error occurs because one letter gets confused with another similar-looking one, both of them compete for the confidence score and consequently the value assigned to the resulting letter is lower than the values for well-recognized letters. Fig. 1 shows an example: The two lowest confidence values are actual errors ($O \to G$ and the insertion of *r*), and other low values correspond to an imperfect recognition (italic *b* and *h* look very similar, *m* is partly recognized as *r*). More importantly, all characters with a confidence above average (0.93) are correct. The sum of the confidences over all output characters should therefore correlate with the accuracy of the output. Systematic problems for this measure arise from deletion errors (e.g., missed blanks between tokens), because deletions by their very definition do not have a confidence value attached to them.

Below we identify the best model among all the models saved during a training history (every 1,000 learning steps) by choosing the one with the best score (confidence or lexicality). This will work as long as there is a good correlation between these scores and accuracy. The exact relation may be different for each document and even for different training methods on the same document. As soon as some ground truth is available for testing, we can compare the different methods for their actual accuracy. Also, from the statistical properties of the correlation one can give prediction intervals at level $\alpha$ for the accuracy of the OCR result of a complete document based on its measured score $x_0$ according to the formula:

$$\hat{y} \pm t_{\alpha/2, n-2} S_y \sqrt{1 + \frac{1}{n} + \frac{n(x_0 - \bar{x})^2}{n \sum x_i^2 - (\sum x_i)^2}} \quad (1)$$

---

[7] The code of OCRopus had to be slightly adapted to output the confidence value of each character.

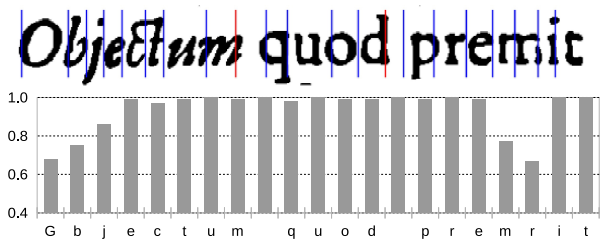

**Fig. 1** Confidence values for characters. Shown is the output of OCRopus using a trained model for Hobbes' Leviathan (1668) for a snippet of a few words (upper image: printed text together with approximate LSTM character boundaries; lower image: recognized characters with their associated confidences.)

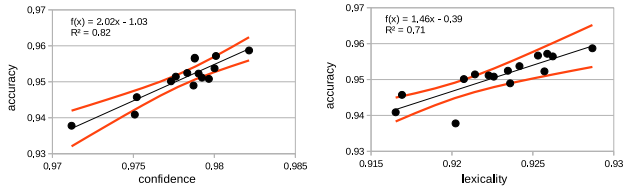

**Fig. 2** Predictors for accuracy for document 1483-D. Left image: Average character confidence of the OCR output versus true accuracy; right image: average lexicality versus accuracy. The black dots represent the OCR output of different models during the training phase. Also shown are the regression line (black) and upper and lower limits (red) of prediction intervals at 95% statistical confidence.

Here, the $(1-\alpha)$-percentile with $(n-2)$ degrees of freedom of Student's T distribution is given by $t_{\alpha/2, n-2}$, and the residual standard error

$$S_y = \sqrt{\frac{\sum(y_i - \hat{y}_i)^2}{n-2}} \quad (2)$$

is calculated from the distances of the $y_i$ from the regression equation $\hat{y} = f(x) = mx + b$ with $n$ data points $(x_i, y_i)$. Fig. 2 shows as an example the correlation of both confidence and lexicality for 1483-D trained on 42 printed lines together with their 95% prediction intervals (red lines). Each point corresponds to one model of the training history. We can therefore say that a mean character confidence of 98% for the OCR result leads to an accuracy interval of (95.26%, 95.69%) with 95% probability.

Figure 3 shows another possible application of confidence values for visualizing difficult lines and pages of a document. For each line we show the average confidence. It is easy to see that lines with low confidence exactly occur at each page break: these lines are the short first line (capitalized running head) and the last line of a page containing just the catchword (the first word or syllable of the next page) providing little context for the neural-network-based recognition. If only one character is wrongly recognized, it has a large effect on the average character confidence for this line. The reduced confidences in pages 11-15 arise from lines printed in italics which were underrepresented in the mixed trained model because of their relative rarity.

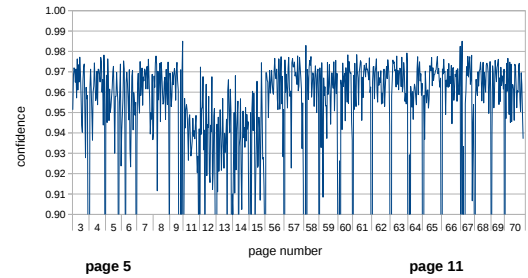

**Fig. 3** Visual inspection of OCR quality by a line/page based *seismographic* view of average character confidence per line for a mixed model result of document 1591-P. The effect of short lines around page breaks (very low average confidences) and a change in typography (italics at pages 11-15) are clearly visible.

## 5 Fully automatic methods for improving OCR on a given document

Our fully automatic procedure for improving OCR on a specific document uses two steps.

*1. Automatic selection of pseudo ground truth using standardized mixed models.* Starting with our standard mixed OCR models (cf. Section 3) we recognized the given document. Two automatic methods are used to define a collection of *pseudo ground truth* (PGT) lines for training. The first method simply takes the full OCR output as a PGT set. The second method is more complex. Using the information provided by the profiler for each token of the OCR output of the initial mixed model, we looked at tokens where the profiler suggests a correction of certain symbols and at the same time the OCR has low confidence for these symbols. We took all lines containing such a token and replaced the original OCR result by the correction suggestion of the profiler.

*2. Automatic training, model evaluation, and new model selection.* Using the two types of PGT we started two training runs with OCRopus on the given document. In each run, new OCR models are saved by OCRopus every 1,000 learning steps. As a result of this automatic training process(es), several alternative OCR models from two runs are at our dis-posal, the start model representing one option. Using the average confidence value of all symbols in the OCR output for each model for the given document as a score we chose the OCR model with the best score. The process is illustrated in Figure 4. The x axis gives the sequence number of the models generated during training. Accuracy (orange), confidence (yellow), and lexicality (brown) show a clear correlation.

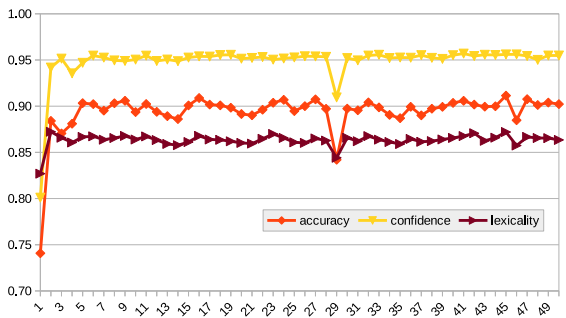

**Fig. 4** Accuracy (orange), confidence (yellow), and lexicality (brown) of a series of models generated during a training run on PGT for 1483-D.

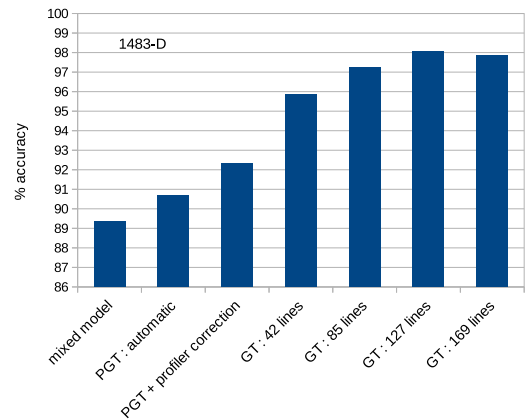

**Fig. 5** Comparison of methods. For document 1483-D, the process of model improvement starting from the mixed (other) model is shown: automatic method with pseudo ground truth (PGT) without and with additional automatic profiler correction, and models trained on an increasing number of manually corrected lines as ground truth (GT). For each method, model selection was based on confidence score and measured for accuracy against available ground truth in our test set.

*Remarks.* Note that for the second step any method for automatic quality evaluation could be used. E.g., profiler info (lexicality) might help when using an OCR that does not offer good confidence values. The two steps can be considered as the first round of an Expectation Maximization (EM) algorithm - the OCR output of the mixed model represents a first kind of expected result, and the training is a first optimization step. We could of course iterate both steps, but we did not follow this direction here.

*Evaluation of the model obtained.* To test the quality of the selected model we used the real ground truth data. Recall that for evaluation purposes full ground truth for a significant part of the documents was at our disposal. Using the selected OCR model we process the document and evaluate the accuracy on this part of the document. Fig. 5 shows the result of this evaluation for document 1483-D with the two types of automatic improvements together with semi-automatic methods involving a number of hand-corrected lines (see Section 6). The automatic methods do improve the initial model but only by a few percentage points, a result that is consistent over the complete data set (see Fig. 6). As long as the initial mixed model gets a chance to recognize the most frequent glyphs in a printing, our two automatic methods lead to OCR results over 90% accuracy for all documents in our data set except in the case of 1476-S, for which later training on PGT cannot remedy the loss of special glyphs that have not been recognized by the mixed model.

The manual correction of a few lines is much more effective and the asymptotic behaviour of adding more training material is visible after about 100 lines of ground truth (GT) have been used. We also checked if better results could be obtained (in theory) when selecting the optimal (highest accuracy) model from the training processes as opposed to the model with the best confidence score. Differences are minor, however, and can be seen in Fig. 5: Whereas for 127 lines of GT the model with best confidence also has the highest accuracy, for 169 lines a slightly inferior model with accuracy of 0.979 instead of 0.981 got selected. As it happens, this model is even slightly worse than the previous result from 127 lines (0.980), but this just shows that we reached a plateau where additonal training material does not immediately lead to clearly discernible progress. Training on 800 lines eventually does give a better result of 0.988, but one has to balance the greatly increased effort of preparing an additional 631 lines of ground truth against any postcorrection activities that could lead to similar or even better results.

## 6 Semi-automatic methods for improving OCR on a given document

The above results with fully automatic improvement indicate that we can achieve *good* (> 90% accuracy) OCR quality, but in order to achieve *excellent* (> 95% accuracy) OCR results for a document a certain amount of manual work seems to be inevitable.

We now ask how to obtain the maximal benefit from a minimal amount of manual work. In our case the manual work only consists in the the simple task to transcribe a small number of text lines of the given input document as a ground truth set for OCR training. Technically, the ground truth for the lines can be prepared using the OCR output for the mixed model and postcorrecting the selected lines using a postcorrection system (Vobl et al 2014).

To optimize the benefit (OCR improvement) obtained from transcribing a fixed number of lines and to minimize manual work we looked at different strategies for line selection: (i) selection of a set of consecutive lines, (ii) selection of a random set of lines from the full document, (iii) selection of a collection of lines with high OCR confidence, (iv) selection of lines with low OCR confidence, and (v) mixtures of these strategies. As a first result worth to be mentioned we found that optimal results are obtained when using a mixture of randomly selected lines plus lines with low OCR confidence. A possible explanation is the following: First, random selection of lines have the effect that many distinct pages and positions are taken into account, which is important to obtain improvements on all parts of a document. Second, assuming the lines with low confidence often have many errors, preparing GT for lines with low confidence optimizes the number of positions where model training leads to a real improvement in OCR recognition. After the selection of the GT material for training, the other steps (training, automatic model selection and evaluation of the selected model) are as above.

In two series of experiments, for each document, using the line selection strategy described above we automatically selected 100 lines (in three cases additionally 200 lines) and trained OCRopus with the GT for this selected set of lines. As a starting point we always used the standard mixed OCR model described above.

Fig. 6 shows the result for both parts of our data set: Accuracies of individually trained models (blue), the mixed (own) model trained from all books contained in this part (orange), the mixed (other) model trained from the books contained in the other part (yellow), the fully automatic improvement method (without profiler intervention; green) and the models trained from a number of manually corrected lines (brown and light blue, respectively). The general picture emerging from these result is that individual models trained from all available ground truth are best (from 94% to 99%), mixed (own) models are slightly worse, mixed (other) models even lower in accuray but are (with the exception of 1476-S with its peculiar type) still in the range of 88% to 98%. The difference between individual and mixed (other) model performance is the improvement potential and the subsequent models gain in that respect, with 1 to 2 percentage points of automatic improvements and a sometimes large jump with models trained on 100 lines of GT, bringing it near the results of the mixed (own) model. For 1533-C, 1471-O and 1522-D the 100-line models are even on a par with the individual models (the differences are within the range of statistical fluctutations and not significant), which is no surprise as these documents have a small amount of training data (see Table 1). Only in three cases (1686-L, 1591-P and 1668-L) are these 100-line results lower than the results of the mixed (other) model and do not represent any improvement at all; however, doubling the number of gt lines to 200 solved this problem. As a guide whether a certain amount of manually corrected lines leads to a trained model with better results than the mixed model one can again use our quality measures of average confidence and lexicality. Also, once additional ground truth for a new book becomes available, the mixed model can be updated to include this book in its training pool, making it more widely applicable to other books to be recognized. In this way a gradual shift from mixed (other) to mixed (own) model results in book recognition can be anticipated.

As the results with models based on 100-200 lines are within a distance of a few percentage points from individual models trained on a sometimes much larger training set, further improvement may be reached more efficiently with postcorrection (see Vobl et al 2014) than with greatly enlarged training sets.

## 7 Summary

In this paper we looked at strategies that help to obtain optimal OCR results on historical documents with a minimal amount of manual work.

Summing up, we suggest to use a set of standard mixed models for OCRopus, each covering a spectrum of periods and printings, as a starting point. Standard models could be prepared and exchanged by the community. Once we have such a set, to process a new book we may use an automatic quality measure such as confidence or lexicality (Sect. 4) to determine the model that offers the best starting point. We may then improve the start model for the given document either in a fully automatic way or by preparing ground truth for a small number of lines.

For finding the best model of the subsequent training runs we again use automatic quality estimation. Quality estimation will also get used to compare the results of the start model and the chosen best automatic or semi-automatic model from later trainings. In this way one can decide whether a specific number of ground truth lines is sufficient for an improved recognition or needs to be expanded. The results in this paper show that in this way really excellent results can be achieved with a minimal amount of manual work. Also, any additional ground truth prepared using the semi-automatic method can be used to update the mixed model, leading to in incremental move from a mixed (other) model to an omnifont mixed (own) model with better recognition results to start from.

As a matter of fact, more data and experiments are needed to make this picture more complete and robust. A second important point for future work is to investigate the correlation between confidence or lexicality and accuracy across distinct documents and OCR models.

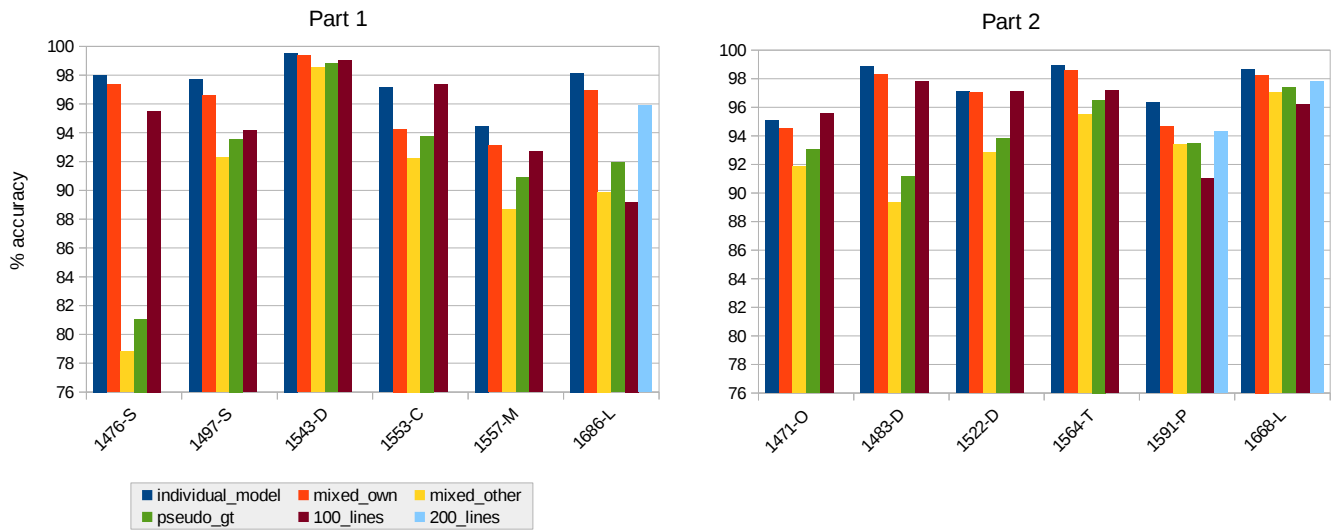

**Fig. 6** Accuracies reached by differently trained models on all documents of our data set: Individual models (blue), mixed (own) models (orange), mixed (other) models (yellow), and subsequent improvements by the fully automatic method based on taking the mixed (other) model result (green) as pseudo ground truth, plus models trained on a number (100 = brown, 200 = light blue) of manually corrected lines.

**Acknowledgements** We wish to thank our students Haide Friedrich-Salgado and Jasmin Chebib for creating the ground truth for Hobbes. The material on Tortellius was provided by our Italian colleagues Federico Boschetti, Paola Tomé and Edoardo Bighin, and the ground truth for Brant has been generously shared by the Kallimachos Project at Universität Würzburg (Hans-Günter Schmidt, Felix Kirchner, and Marco Dittrich).